\documentclass[aps,prd,english,superscriptaddress,11pt,notitlepage]{revtex4}
	\usepackage[colorlinks=true, a4paper=true, pdfstartview=FitV,
linkcolor=blue, citecolor=blue, urlcolor=blue]{hyperref}
\usepackage{amsmath}
\usepackage{amssymb}
\usepackage{graphicx}
\usepackage{hhline}
\usepackage{textcomp}
\makeatletter
\usepackage{babel}
\newcommand{\bea}{\begin{eqnarray}}
\newcommand{\eea}{\end{eqnarray}}

\newcommand{\be}{\begin{equation}}
\newcommand{\ee}{\end{equation}}
\usepackage[active]{srcltx}
\begin{document}

%\immediate\write16{<<WARNING: LINEDRAW macros work with emTeX-dvivers
%                    and other drivers supporting emTeX \special's
%                    (dviscr, dvihplj, dvidot, dvips, dviwin, etc.) >>}

 \title{Incompressible topological solitons}
 
 \author{C. Adam}
\affiliation{Departamento de F\'isica de Part\'iculas, Universidad de Santiago de Compostela and Instituto Galego de F\'isica de Altas Enerxias (IGFAE) E-15782 Santiago de Compostela, Spain}
\author{C. Naya}
\affiliation{INFN, Sezione di Lecce, Via per Arnesano, C. P. 193 I-73100 Lecce, Italy}
\author{K. Oles}
\affiliation{Institute of Theoretical Physics,  Jagiellonian University, Lojasiewicza 11, Krak\'{o}w, Poland}
\author{T. Romanczukiewicz}
\affiliation{Institute of Theoretical Physics,  Jagiellonian University, Lojasiewicza 11, Krak\'{o}w, Poland}
\author{J. Sanchez-Guillen}
\affiliation{Departamento de F\'isica de Part\'iculas, Universidad de Santiago de Compostela and Instituto Galego de F\'isica de Altas Enerxias (IGFAE) E-15782 Santiago de Compostela, Spain}
\author{A. Wereszczynski}
\affiliation{Institute of Theoretical Physics,  Jagiellonian University, Lojasiewicza 11, Krak\'{o}w, Poland}

\begin{abstract}
We discover a new class of topological solitons. These solitons can exist in a space of infinite volume like, e.g., $\mathbb{R}^n$, but they cannot be placed in any finite volume, because the resulting formal solutions have infinite energy. These objects are, therefore, interpreted as totally incompressible solitons. 

As a first particular example we consider (1+1) dimensional kinks in theories with a nonstandard kinetic term or, equivalently, in models with the so-called runaway (or vacummless) potentials. But incompressible solitons exist also in higher dimensions. As specific examples in (3+1) dimensions we study Skyrmions in the dielectric extensions both of the minimal and the BPS Skyrme models. In the the latter case, the skyrmionic matter describes a completely incompressible topological perfect fluid. 
\end{abstract}

\maketitle
%%%%%%%%%%%%%%%%%%%%%%%%%%%%%
\section{Introduction}
%%%%%%%%%%%%%%%%%%%%%%%%%%%%%
Topological solitons are ubiquitous objects in modern physics, both from a theoretical point of view and in a variety of applications \cite{manton2004topological}, \cite{shnir2018topological}. They are particle-like solutions of non-linear field theories characterized by a pertinent topological index (charge) $Q$, whose conservation is not related to the Noether theorem, but is a consequence of some topological properties of the physical (base) space $\mathcal{M}$ and the field space of the theory (target space $\Sigma$).

The stability of topological solitons is guaranteed by the existence of the so-called topological energy bound, which states that the energy $E$ of any field configuration is bounded from below by the topological degree. Typically, the bound takes a linear form \cite{manton2004topological}
\be
E \geq C |Q|, \label{bound-1}
\ee
 although theories with nonlinear versions are also known \cite{vakulenko1979stability}, \cite{aratyn1999exact}. Here, $C$ is a numerical constant which does not depend on the volume of the base space. Hence, this bound applies to infinite (e.g. $\mathcal{M}=\mathbb{R}^n$) as well as to finite volume base spaces. In some very special theories, the bound can be saturated, which gives rise to Bogomolny-Prasad-Sommerfield (BPS) solitons \cite{bogomolny1976stability}. They satisfy lower order field equations (which obviously imply the usual Euler-Lagrange equations) called BPS or self-dual (SD) equations and are, therefore, mathematically much simpler, often allowing for an analytical treatment. Physically, BPS solutions explore the limit where static solitons do not interact, which results in zero binding energies. 

If a soliton is considered on a finite volume manifold, $\mbox{vol} (\mathcal{M}) < \infty$, then, frequently, another topological energy bound can be derived, 
\be
E \geq C_{\mbox{vol}(\mathcal{M})} f(|Q|). \label{bound-2}
\ee 
In contrast to the BPS bound mentioned above, this second bound usually is not linear in $Q$  \cite{manton1987geometry}, \cite{harland2014topological}, \cite{adam2014topological}. Furthermore, $C_{\mbox{vol}(\mathcal{M})}$ is a function of the volume of the base space. Thus, this new bound 
does depend on the volume. The two bounds are, of course, independent. Hence, for some values of the model parameters and/or topological charges, one of them provides a tighter bound. Physically, the volume dependent bound encodes some information about the resistance of the soliton against external pressure. Indeed, it shows how the energy grows if a soliton is forced to occupy a finite volume space $V=\mbox{vol} (\mathcal{M})$. This leads to a very important quantity characterizing a soliton, which is its compressibility $\kappa$ defined as
\be
\kappa=-\frac{1}{V} \left( \frac{\partial V}{\partial p} \right)_{Q, T}
\ee
where $V$ is the volume of the soliton, $p$ is its pressure and $T$ the temperature. 

If solitons are classified according to their size (occupied volume), then, currently, there are two known types:
\begin{itemize}
\item {\it Usual solitons}, which at zero pressure are infinitely extended solutions approaching the vacuum at $\vert \vec{x} \vert \to \infty$. Obviously, they possess infinite volume $V$.
\item {\it Compactons},  i.e., solitons which even at zero pressure approach their vacuum values at a finite distance and, therefore, have a finite volume $V$, see e.g., \cite{Rosenau:1993zz, Arodz:2002yt, Arodz:2005gz, Adam:2007ij, Bazeia:2008tj, Arodz:2008nm, Speight:2010sy, Hartmann:2013kna, Kumar:2019dbi, Gudnason:2020tps}.
\end{itemize}
When a non-zero pressure is applied, solitons of both types reduce their volumes. Obviously, it requires additional  energy to keep the solitons in the reduced volume. However, this energy is always finite although it may rise quickly as $V$  decreases (or, equivalently, as $p$ increases). Therefore, all known solitons have non-zero compressibility and can be squeezed to smaller sizes with a finite amount of energy. 

It is the aim of the current paper to prove the existence of {\it a new class of topological solitons} which, although they exist in an infinite volume space, e.g., in $\mathcal{M}=\mathbb{R}^n$,  cannot be squeezed to a finite volume, which means that their compressibility is zero. This possibility can be understood from the independence of the two topological bounds. Indeed, as we will show below, it may happen that for a given solitonic model the constant $C$ is finite while $C_{\mbox{vol}(\mathcal{M})}=\infty$, which prevents the existence of finite energy solutions with non-trivial values of the topological charge in a finite space. In a sense, this new class is exactly opposite to compactons, which even without pressure are finite volume objects. Therefore, it  provides the second extreme limit for the possible qualitative behavior of topological solitons.

\vspace*{0.2cm}

For simplicity, we start with incompressible kinks in (1+1) dimensions (Sec. II). The examples of incompressible kinks will be found in scalar models with a nonstandard kinetic term. Interestingly, by a field transformation, these models can be recast into theories with a standard kinetic term but with potentials belonging to the so-called runaway (vacuumless) class (Sec. III). Then we show that incompressible topological solitons can exist in higher dimensions, as well. Concretely, we consider two examples provided by the recently introduced dielectric generalizations of the minimal Skyrme model (Sec. IV) and BPS Skyrme model (Sec. V). Especially the latter case, which describes an incompressible perfect fluid solitonic matter, allows us to fully clarify the reasons that forbid the existence of finite energy solutions when an external pressure is applied. Although in all examples we deal with BPS theories, this is by no means a necessary condition to find an incompressible soliton. However, it simplifies computations and permits an analytical treatment. 

 %%%%%%%%%%%%%%%%%%%%%%%%%%%%%%%%%%
 \section{Incompressible kinks}
 %%%%%%%%%%%%%%%%%%%%%%%%%%%%%%%%%%
  %%%%%%%%%%%%%%%%%%%%%%%%%%%%%%%%%%
 \subsection{The Bogomolny sector}
 %%%%%%%%%%%%%%%%%%%%%%%%%%%%%%%%%%

 We consider a real scalar field theory in (1+1) dimension with a nonstandard kinetic term, $\Sigma=\mathcal{M}=\mathbb{R}$. Specifically, we promote the coupling constant in front of the kinetic term to a field (target space) dependent function $g(\phi)$
 \be
 L=\int_{-\infty}^\infty dx \left( g(\phi) (\partial_{\mu} \phi)^2 - V(\phi)\right). \label{gen-kinks}
 \ee
 We assume that the potential has two isolated vacua, $\phi_+>\phi_-$, which are attained in a quadratic manner. It means that for  field values close to the vacuum values, $\phi= \phi_\pm - \zeta +O(\zeta^2)$, the potential is $V(\phi) = \frac{1}{2} V''(\phi_\pm) \zeta^2 + O(\zeta^3)$. For our purposes, it is essential that the coupling function $g$ has poles exactly at the same points where the zeros of the potential $V$ are located. Therefore, for simplicity, we restrict our considerations to the case where $g(\phi)=1/V(\phi)$ (a generalization to $g(\phi)=f(\phi)/V(\phi)$ where $f(\phi)$ is a smooth function without zeros and poles is straightforward) and arrive at the following theory
 \be
 L=\int_{-\infty}^\infty dx \left(\frac{1}{V(\phi)} (\partial_{\mu} \phi)^2 - V(\phi)\right). \label{uncomp-kink}
 \ee 
 We comment that models (\ref{gen-kinks}) are widely considered in the literature, although the particular properties of the class (\ref{uncomp-kink}) which we want to discuss have not been noticed yet. 
 
The static energy can be bounded from below by the standard Bogomolny trick, 
\bea
 E&=&\int_{-\infty}^\infty dx \left(\frac{1}{V(\phi)} \phi_x^2 + V(\phi)\right) = \int_{-\infty}^\infty dx \left(\frac{1}{\sqrt{V(\phi)}} \phi_x \pm \sqrt{V(\phi)} \right)  \mp 2 \int_{-\infty}^\infty dx \phi_x \nonumber \\
 &\geq & 2 \left| \int_{-\infty}^\infty \phi_x dx \right| = 2(\phi_+-\phi_-) |Q| \label{1d-bound}
\eea
where $Q$ is a topological charge normalized to $\pm 1$, 
\be
Q= \frac{1}{\phi_+-\phi_-}  \int_{-\infty}^\infty \phi_x dx = \frac{\phi(\infty) - \phi(-\infty)}{\phi_+-\phi_-}.
\ee
The bound is saturated if and only if the following Bogomolny (BPS) equation is obeyed
\be
\frac{1}{\sqrt{V(\phi)}} \phi_x \pm \sqrt{V(\phi)} =0 \;\; \Rightarrow \;\; \phi_x = \mp V(\phi).
\ee
Obviously, the BPS equations give rise to kink and antikink BPS solutions. 

In general, the BPS sector is completely standard and is, in fact, identical to the standard kink model with potential $V^2$. Interestingly, this is no longer the case for non-BPS solutions. 
 %%%%%%%%%%%%%%%%%%%%%%%%%%%%%%%%%%
 \subsection{Constant pressure solutions}
 %%%%%%%%%%%%%%%%%%%%%%%%%%%%%%%%%%
Formally, the full static second order Euler-Lagrange equation can be integrated to the following one-parameter equation, which is a constant pressure generalization of the BPS equation,
\be
\frac{1}{V(\phi)}  \phi_x^2 - V(\phi)=P.
\ee
Here $P$ is a constant which can be easily identified with the $T_{11}$ component of the energy-momentum tensor. 
Indeed, if we differentiate it with respect to $x$ we get
\be
\frac{2}{V} \phi_{xx} - \frac{V_\phi}{V^2} \phi_x^2 -V_\phi=0
\ee
which is exactly the static EL equation. The constant pressure equation, being a first order ordinary differential equation, allows to change the base space "volume" measure to the target space measure (we choose the plus sign)
\be
dx=  \frac{d\phi}{\sqrt{V^2+PV}}. \label{dx}
\ee
Therefore, the static energy functional can be rewritten as
\be
E=\int_{-\infty}^\infty dx \left(\frac{1}{V(\phi)} \phi_x^2 + V(\phi)\right) = \int_{\phi_-}^{\phi_+} d\phi \frac{2V+P}{\sqrt{V^2+PV}}
\ee
which is just a target space integral. For $P=0$ we get $E=2\int_{\phi_-}^{\phi_+} d\phi$ which gives a finite result coinciding with our previous expression (\ref{1d-bound}). On the other hand, for $P>0$, there is a divergence at the vacuum. In fact, let us consider the limit when $\phi \to \phi_+$
\be
 \lim_{\phi \to \phi_+} \int^{\phi} d\phi' \frac{2V+P}{\sqrt{V(V+P)}} \sim \lim_{\phi \to \phi_+} \int^{\phi} d\phi' \frac{P}{\sqrt{VP}} =  \sqrt{P} \lim_{\phi \to \phi_+} \int^{\phi} d\phi' \frac{1}{\sqrt{V}} 
\ee
where the last integral diverges if the approach to the vacuum is quadratic (or stronger), which we previously assumed for the potential. The conclusion is that in this case the formal constant pressure solution possesses infinite energy. 

Surprisingly, contrary to usual solitons and compactons, the configurations with $P>0$ still extend to infinity. To see this, we integrate (\ref{dx}). Then,
\be
V = \int dx = \int_{\phi_-}^{\phi_+} \frac{d\phi}{\sqrt{V(V+P)}}
\ee
which diverges at the vacua for any positive pressure. Hence, the volume of solitons of this new type remains infinite despite the application of a nonzero pressure. This means that a constant pressure is not sufficient to compress the solitons to a finite domain. As we will show later, these features are shared by incompressible solitons also in higher dimensions. 

To see the impact of pressure on the infinite energy solutions, we analyze the constant pressure equation in the limit close to the vacuum. We start with $P=0$ and consider the asymptotic behavior at $x \to \infty$, where the field approaches the larger vacuum value $\phi_+$ (the approach to the smaller vacuum is analogous). Here, $\phi=\phi_+ - \zeta +O(\zeta^2)$, where $\zeta$ obeys
\be
\frac{\zeta_x^2}{\zeta^2}= \left( \frac{V''(\phi_+)}{2} \right)^2\zeta^2.
\ee
As a consequence, the decay of the field is power-like, $\zeta \sim x^{-1}$, which is fast enough to guarantee the finiteness of the energy. Now, for non-zero pressure $P>0$, close to the vacuum the behavior changes. Indeed, the asymptotic field $\zeta$ obeys
\be
\frac{\zeta_x^2}{\zeta^2}  = \frac{V''(\phi_+)}{2} P,
\ee
as the contribution from the potential term can be neglected. Therefore, the constant pressure solutions are exponentially localized, $\zeta \sim e^{-\sqrt{\frac{P V''(\phi_+)}{2}}x}$. Hence, the nonzero pressure leads to a better localization of the kink but, simultaneously, results in an asymptotically constant kinetic term $\phi^2_x/V \sim \zeta_x^2/\zeta^2 \sim \mbox{const.}$, which is the origin of the divergency of the energy integral. 

On the other hand, the static BPS solution can be perturbed by any local deformation provided that it decreases in a power like manner, i.e., $\zeta \sim x^{-a}$, $a>0$. Then, the kinetic terms decreases as $x^{-2}$ which is obviously integrated to a finite number. This counterintuitive feature, that a better localization of the field results in worse convergence of the energy integral, is obviously a direct result of the nontrivial kinetic term. 

\vspace*{0.2cm}

Some further intuition concerning a near vacuum perturbation can be achieved if we regularize the kinetic term by considering the following Lagrangian 
 \be
 L_\epsilon=\int_{-\infty}^\infty dx \left(\frac{1}{V(\phi)+\epsilon} (\partial_{\mu} \phi)^2 - V(\phi)\right). \label{reg-kink}
 \ee 
 where $\epsilon$ is a small parameter which finally should be taken to 0. For potentials with a quadratic near vacuum approach it gives a standard Lagrangian for a small perturbation $\zeta$ 
 \be
 L_\epsilon[\zeta]=\int_{-\infty}^\infty dx \left(\frac{1}{\epsilon} (\partial_{\mu} \zeta)^2 - \frac{1}{2}V''(\phi_\pm) \zeta^2 \right). \label{reg-kink}
 \ee
 where terms up to $\zeta^2$ are kept. Thus, the mass of the small linear perturbation is 
 \be
 m_{\epsilon} = \frac{\epsilon V''(\phi_\pm)}{4}
 \ee
 and goes to 0 as we approach the original theory, i.e., $\epsilon \to 0$. Note that the analogous regularization for compactons provides an infinite mass of small (linear) perturbations. Hence, our solitons are, in a sense, exactly opposite to the compacton limit. 

 %%%%%%%%%%%%%%%%%%%%%%%%%%%%%
\subsection{Nonexistence of finite volume kinks}
%%%%%%%%%%%%%%%%%%%%%%%%%%%%%
The fact that the finite pressure solutions have infinite energy and are infinitely extended does not necessary imply that there are no finite volume solutions for the kinks considered here. However, it is not difficult to show that any finite volume topologically nontrivial solution of (\ref{uncomp-kink}) must have infinite energy. This is the place where another, finite volume, topological bound enters. 

To prove it, we use a version of the H\"{o}lder inequality
\be
\left( \int_\mathcal{M} \Omega_\mathcal{M} |f|^p \right) \geq \frac{1}{(\mbox{vol}(\mathcal{M}))^{p/q}}\left( \int_\mathcal{M} \Omega_\mathcal{M} |f| \right)^p
\ee
where the positive numbers $p, q$ are such that 
\be 
\frac{1}{p} +\frac{1}{q}=1 
\ee
and $\mbox{vol} (\mathcal{M})$ is the volume of the base space. Now, the static energy can be bounded by another topological bound
\bea
 E&=&\int_{-\infty}^\infty dx \left(\frac{1}{V(\phi)} \phi_x^2 + V(\phi)\right) = \int_{-\infty}^\infty  \frac{1}{V(\phi)} \phi_x^2 dx  =  \int_{-\infty}^\infty \left(  \frac{1}{\sqrt{V(\phi)}} \phi_x \right)^2 dx  \nonumber \\
 &\geq & \frac{1}{\mbox{vol} (\mathcal{M})} \left( \int_{-\infty}^\infty \frac{1}{\sqrt{V(\phi)}} \phi_x  dx \right)^2 =  \frac{1}{\mbox{vol} (\mathcal{M})}   \left( \int_{\phi_-}^{\phi_+} \frac{d\phi}{\sqrt{V(\phi)}}  \right)^2 =\infty
\eea
where the last integral takes an infinite value due to the logarithmic divergency. 

Hence, the BPS (anti)kinks, although they exist on $\mathbb{R}$, cannot be squeezed to a finite volume. Thus, the kink and antikink of the nonstandard kinetic term model presented above are examples of incompressible solitons.

%%%%%%%%%%%%%%%%%%%%%%%%%%%%%
\section{Formulation as a runaway potential model}
%%%%%%%%%%%%%%%%%%%%%%%%%%%%%
%%%%%%%%%%%%%%%%%%%%%%%%%%%%%
\subsection{Incompressible kinks in runaway potential models}
%%%%%%%%%%%%%%%%%%%%%%%%%%%%%
The coupling function $g(\phi)=1/V(\phi)$ in the model supporting incompressible kinks (\ref{uncomp-kink}) can be viewed as a nontrivial metric on one dimensional target space $\Sigma$. Due to its one-dimensionality such a metric can always be made locally trivial by a suitable field redefinition, $\phi=\phi(\psi)$,
\be
\frac{d\phi}{\sqrt{V}} = d\psi
\ee
which leads to a a scalar field theory with the canonical kinetic term
\be
L[\psi] =\int_{-\infty}^\infty dx \;  (\partial_\mu \psi)^2 - \tilde{V}(\psi).  \label{vacuumless}
\ee
(This target space transformation has been very recently used in the context of domain walls without a potential \cite{Deffayet:2020vrh}.) Of course, the form of the potential $\tilde{V}(\psi)\equiv V(\phi(\psi))$ changes. The characteristic feature of the potential in the variable $\psi$, i.e., $\tilde{V}(\psi)$, is that its vacua $\tilde{V}(\psi)\to 0$ are approached  in the limit $\psi=\pm \infty$. This is an obvious consequence of the formula relating the fields. Indeed, as $V(\phi)$ has at least a quadratic approach to the vacuum, $\phi \to \phi_\pm$ leads to $\psi \to \psi_\pm = \pm \infty$. Such potentials are called  {\it vacuumless} or {\it runaway} potentials and have been widely considered in the literature \cite{cho_vacdef, cho_grav, Bazeia1999, Morris2003runaway, Chiba:2011en, Bamba:2011nm}. The first name might be considered a bit misleading as the potential still approaches two vacua, although in the limit $\psi \to\pm \infty$. Therefore, we will use the second name. These potentials still support BPS topological solitons interpolating between the infinitely separated vacua $\psi_\pm = \pm \infty$. The pertinent Bogomolny equation is
\be
\psi_x = \pm \sqrt{\tilde{V}(\psi)}
\ee
with solutions (kink and antikink) saturating the topological energy bound
\be \label{E-bound-psi}
E[\psi] \geq 2 \int_{\psi_-}^{\psi_+} \sqrt{\tilde{V}(\psi)} d\psi = 2 \int_{-\infty}^{\infty} \sqrt{\tilde{V}(\psi)} d\psi .
\ee
Observe that the target space integral is over an infinite volume space $\Sigma=\mathbb{R}$ and, therefore, its convergence requires a sufficiently fast approach to the vacua. However, due to the equivalence of the nonstandard kinetic model (\ref{uncomp-kink}) and the runaway theory (\ref{vacuumless}), the integral takes a finite value. In general, at the vicinity of the vacuum the approach should be at least $\tilde{V}(\psi) \approx \psi^{-a}$, with $a>2$, or faster. 

Furthermore, all results concerning the existence of incompressible kinks hold in the runaway potential model (\ref{vacuumless}). Therefore, such theories also support incompressible solitons. Let us for example consider the constant pressure equation
\be
\psi_x^2  = P +\tilde{V}(\psi). 
\ee
For topologically nontrivial solutions it is necessary for the field to approach the vacua $\psi \to \pm \infty$. Then, the potential vanishes which close to the vacuum gives $\psi_x^2 = P$. This leads to a linear divergency of the field at spatial infinities, i.e., $\psi \sim x $ as $x\to \pm \infty$. But this results in a divergency of the kinetic part of the total energy. So, exactly as in the case of the nonstandard kinetic term, constant pressure solutions are formal solutions possessing infinite energy. 

To clearly understand this equivalent formulation we consider a particular example, which is the $\phi^4$ potential, 
\be
V_{\phi^4}=(1-\phi^2)^2.
\ee
The incompressible kinks of the nonstandard kinetic term model (\ref{uncomp-kink}) are given in an implicit form
\be
\frac{1}{4}\left(\frac{2\phi}{1-\phi^2}  - \ln \left| \frac{1+\phi}{1-\phi} \right| \right) = \pm (x-x_0)
\ee
where $x_0$ is a free parameter i.e., the location of the (anti)kink. The pertinent change of the field is $\phi = \tanh \psi$. This map relates $\phi \in [-1,1]$ with $\psi \in \mathbb{R}$. The resulting potential is
\be
\tilde{V}(\psi)= \frac{1}{\cosh^4 \psi},
\ee
while the topological (anti)kinks in the variable $\psi$ read
\be
\frac{1}{2} \left( \psi + \frac{1}{2} \sinh (2\psi) \right) = \pm (x-x_0).
\ee
The runaway kinks are examples of solitons with very long tails \cite{Manton:2018deu}, \cite{Christov:2018ecz}. Indeed, the energy density decreases as $1/x^2$. 

We remark that the original fields outside of this segment, i.e., $|\phi | > 1$, can be parameterized as $\phi  = \coth \chi $. This again gives a model with the standard kinetic part but now the potential is 
\be
\tilde{V}(\chi) = \frac{1}{\sinh^4 \chi}.
\ee
This is also a runaway potential with two vacua at $\chi = \pm \infty$. However, there are no kinks interpolating between them. The reason is that the potential has an infinite barrier at $\chi =0$. Nonetheless, the full dynamics of the original model based on the $\phi$ field may require to take into account also this branch. Of course, the runaway model based entirely on the $\psi$ field (\ref{vacuumless}) can be considered as a fully self-consistent dynamical system. In this case, it would correspond to the nonstandard kinetic term model with $\phi \in [-1,1]$. 
%%%%%%%%%%%%%%%%%%%%%%%%%%%%%
\subsection{Mode structure}
%%%%%%%%%%%%%%%%%%%%%%%%%%%%%
Although a full analysis of the dynamical properties of incompressible solitons goes beyond the scope of this paper, we present the main features of the mode structure which describes the behavior of small perturbations around the incompressible kink. Taking into account the equivalence of the models supporting incompressible solitons defined above, we will use only one of them, namely, the runaway potential model (\ref{vacuumless}). 

The common feature of all runaway models is that the mass of small perturbations is zero, $m^2=0$. Indeed, in the expansion of the potential at the vacua there is no term proportional to $\psi^2$. This agrees with our previous comment on the regularized limit of the nonstandard kinetic term. As a consequence, there is {\it no mass gap} in the spectrum of the theory. The mass threshold which divides the discrete and continuous spectrum starts at $\omega^2=m^2=0$. 

Now we deform the incompressible (anti)kink by a small perturbation $\eta(x,t)$. Inserting $\psi(x,t)= \psi_{kink} (x) + \eta(x,t)$, where $\eta(x,t)=\eta(x)e^{i\omega t }$, into the Euler-Lagrange equation and leaving only linear terms in the perturbation we get a Schr\"odinger-like equation
\be
-\frac{d^2}{dx} \eta (x) + V_{lin}(x) \eta (x) = \omega^2 \eta
\ee
where the linearized potential is
\be
V_{lin}(x)= \left. \frac{d^2\tilde{V}}{d\psi^2} \right|_{\psi=\psi_{kink}(x)} .
\ee
Normalizable solutions with $\omega \in \mathbb{R}$ are normal modes. 
Here, the only normal mode is the zero mode related to the translational invariance of the model. This mode generates the translations of the free (anti)kink. No other normal modes are possible, because the mass threshold is located at $\omega^2=m^2=0$. It separates the discrete and continuous spectrum. So, there is no room for any other bound mode, while unstable modes $\omega^2<0$ are forbidden by the saturation of the energy bound (\ref{E-bound-psi}). 
\begin{figure}
\center \includegraphics[width=0.72\textwidth]{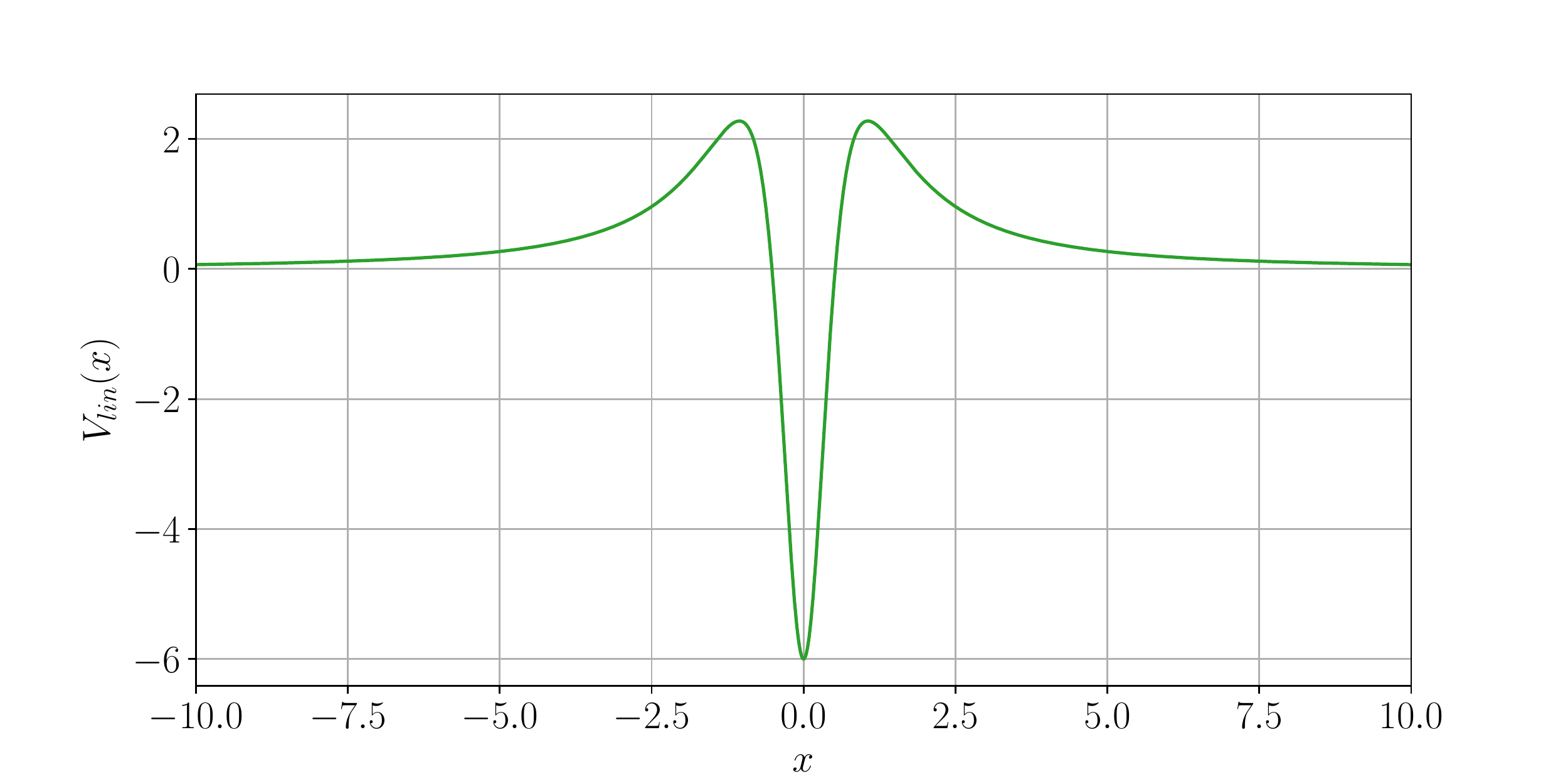}
\caption{Linearized potential in the small perturbation problem for the runaway model with $\tilde{V}=1/\cosh^4\psi$. Note the volcano shape.}
\label{volcano} 
\end{figure}

The nonexistence of massive normal modes for incompressible kinks can make their dynamics relatively simply. The reason is that, e.g., complicated chaotic structures in kink-antikink collisions are mainly related to the existence of a massive bound mode. This is the case for the shape mode in $\phi^4$ theory. Indeed, during the collision initial kinetic energy can be temporarily stored in bound modes and then released in the so-called resonant mechanism which is believed to lead to the fractal structure observed in the final state \cite{Campbell:1983xu}. However, it should be underlined that the existence of a massive normal mode is {\it not} mandatory for the appearance of a resonant structure. Important counterexamples are known, e.g., \cite{Dorey:2011yw}. 

In contrast to massive bound modes, incompressible kinks may possess quasi-normal modes (QNM), which are solutions of the linearized perturbation equation with a complex frequency $\omega=\Omega+i \Gamma$, where both $\Omega$ and $\Gamma$ are real and $\Gamma >0$. Physically, they describe decaying perturbations. 

We find that the runaway potential $\tilde{V}=1/\cosh^4\psi$ leads to a volcano-shaped linearized potential $V_{lin}(x)$,  see Fig. \ref{volcano} (also \cite{Bazeia1999}).  Volcano-shaped potentials tend to support the formation of  QNM, because the potential edges form a sort of barrier which may host oscillating perturbations. Of course, as the barrier is finite the perturbations will eventually decay. The decay rate, $\Gamma$, is smaller if the barrier is higher or wider. In our example, we find that there exist at least three QNM, two for anti-symmetric boundary conditions with frequencies $\omega_1= 0.0077+0.024 i$ and $\omega_3 = 0.43 + 1.27 i$, and one for symmetric boundary conditions with $\omega_2 = 0.023 + 0.012 i$. Note that
at least $\omega_1$ and $\omega_2$ are very low-lying, and their wave functions spread out quite far, so that their existence is probably not related to the volcano shape. Further, although $\Gamma_i \sim \Omega_i$ in all three cases, oscillatory behavior in the QNM is always well visible in our numerics. 

%%%%%%%%%%%%%%%%%%%%%%%%%%%%%
\section{Incompressible Skyrmions}
%%%%%%%%%%%%%%%%%%%%%%%%%%%%%
Now we will show that the phenomenon of incompressible solitons is not confined to one spatial dimension but can occur in higher dimensional solitonic models, as well. Let us consider the so-called dielectric Skyrme model \cite{Adam:2020iye} (see also \cite{naya2020background} and \cite{Gudnason:2020ftf}) which is a variant of the Skyrme model \cite{skyrme1994non}, \cite{skyrme1962unified} with the coupling constants $e$ and $f$ promoted to field-dependent functions. Specifically, in the minimal version it reads 
\be
L^d_{24}=L_2^d+L_4^d=  \int_{\mathcal{M}} \frac{f^2}{2} \mbox{ Tr } (R_\mu R^\mu) d\Omega_\mathcal{M}  - \int_{\mathcal{M}} \frac{1}{16 e^2} \mbox{ Tr } ([R_\mu, R_\nu] [R^\mu,R^\nu]) d\Omega_\mathcal{M} . \label{model}
\ee
Here the Skyrme field is a map $U: \mathcal{M} \longrightarrow \Sigma$, where $\mathcal{M}$ is a three-dimensional manifold without boundary and with volume element $d\Omega_\mathcal{M}$, such that the map $U: \mathcal{M} \longrightarrow \Sigma$ is sufficiently smooth and exists globally.
Further, the target space is just the unit three-dimensional sphere  $\Sigma \equiv \mathbb{S}^3$, and $R_\mu=\partial_\mu U U^{-1}$ is the right invariant current. These maps are classified by a topological index called the baryon charge $Q=B$ defined as
\be
B=\int_{\mathcal{M}} d\Omega_\mathcal{M} \mathcal{B}^0= \frac{1}{24\pi^4} \int_{\mathcal{M}} d\Omega_\mathcal{M}\epsilon^{ijk} \mbox{Tr }(R_iR_jR_k)
\ee
where $\mathcal{B}^0$ is the temporal component of the baryon current $\mathcal{B}^\mu =  \frac{1}{24\pi^4} \epsilon^{\mu \nu \rho \sigma} \mbox{Tr }(R_\nu R_\rho R_\sigma)$. 

If written in terms of the eigenvalues $\lambda^2$ of the strain tensor $D_{i}{}^j=-\frac{1}{2} \mbox{Tr }(R_i R^j)$, the static energy takes the following form \cite{manton1987geometry}
\be
E^d_{24}=\int_{\mathcal{M}} \left[ f^2\left( \lambda^2_1+\lambda^2_2+\lambda^2_3\right) +  \frac{1}{e^2}\left(\lambda^2_1\lambda^2_2+\lambda^2_2\lambda^2_3 + \lambda^2_3\lambda^2_1\right) \right] d\Omega_\mathcal{M} .
\ee
This is bounded from below as 
\be
E^d _{24} \geq  6 \left| \int_{\mathcal{M}} \frac{f}{e}  \lambda_1 \lambda_2 \lambda_3 d\Omega_\mathcal{M} \right| = 12 \pi^2 \left\langle\frac{f}{e} \right\rangle |B|=6\, \mbox{vol}(\Sigma) \left\langle\frac{f}{e} \right\rangle |B|.
\ee
where $\left\langle \mathcal{F} \right\rangle$ is the average value of $\mathcal{F}$ over the whole $\mathbb{S}^3$ target space
\be
\left\langle \mathcal{F} \right\rangle = \int \frac{d\Omega_\Sigma}{2\pi^2} \mathcal{F} =\frac{1}{2\pi^2} \int_0^\pi d\xi \int_0^\pi d\Theta \int_0^{2\pi} d\Phi \sin^2\xi \sin \Theta \; \mathcal{F} (\xi, \Theta, \Phi) .
\ee
Here $(\xi, \Theta, \Phi)$ are coordinates on $\Sigma$ and $d\Omega_\Sigma = \sin^2\xi \sin \Theta d\xi d\Theta d\Phi$ is the volume element. Note that the bound is valid for any sufficiently well-behaved base space manifold $\mathcal{M}$. It is a generalization of the Skyrme-Faddeev bound \cite{faddeev1976some} to the case when the coupling constants are target space functions.

The bound is saturated if and only if
\be
\lambda_1^2=\lambda_2^2=\lambda_3^2= e^2 \left( \mbox{Tr } U \right) f^2 \left( \mbox{Tr } U \right). \label{lambda}
\ee
Contrary to the standard minimal Skyrme model, where the couplings are just constants, this set of equations has a nontrivial $B=1$ solution on $\mathcal{M}=\mathbb{R}^3$ if (here, $r_0$ is a constant with the units of length) 
\be
ef = \frac{1}{2r_0} \mbox{ Tr } (\mathbb{I}-U) \label{cond}
\ee
which we call the BPS constraint \cite{Adam:2020iye}. The pertinent solution is a hedgehog (spherically symmetric) solution 
\be
U=e^{i\xi \vec{n}\cdot \vec{\tau}}, \;\;\; \vec{n}=\frac{\vec{r}}{r}, \;\;\; \xi=2\arctan \frac{r_0}{r}
\ee
where $\vec{\tau}$ are the Pauli matrices while $\xi$ and $\vec{n}=(\sin \Theta \cos \Phi, \sin \Theta \sin \Phi, \cos \Theta)$ are again coordinates on $\Sigma$. Furthermore, for the hedgehog $\Theta = \theta$, $\Phi=\phi$ where $(r, \theta, \phi)$ are the usual spherical polar coordinates. 

For a finite volume manifold $\mathcal{M}$, we can derive another bound which, in some cases, provides a stronger bound on the energy. We remark that finite volume Skyrmions are intimately related to Skyrmionic crystals \cite{manton1987geometry}, \cite{castillejo1989dense}, \cite{kugler1988new}, \cite{perapechka2017crystal}, which, as ground states of the Skyrme model for $B \to \infty$, play a very important role in the application of the Skyrme model to nuclear matter. Here we will use two inequalities. Namely, the arithmetic mean-geometric mean (AM-GM) inequality
\be
\sum_{i=1}^n a_i \geq n \left( \prod_{i=1}^n a_i \right)^\frac{1}{n}
\ee
and the previously used H\"{o}lder inequality. 
Now, we again use the static energy expressed in terms of the eigenvalues
\bea
E^d_{24}&=& \int_{\mathcal{M}} \left[ f^2\left( \lambda^2_1+\lambda^2_2+\lambda^2_3\right) +  \frac{1}{e^2}\left(\lambda^2_1\lambda^2_2+\lambda^2_2\lambda^2_3 + \lambda^2_3\lambda^2_1\right) \right] d\Omega_\mathcal{M} \nonumber \\
&\geq &  \int_{\mathcal{M}}  \frac{1}{e^2}\left(\lambda^2_1\lambda^2_2+\lambda^2_2\lambda^2_3 + \lambda^2_3\lambda^2_1\right) d\Omega_\mathcal{M} \nonumber \\
&\geq &3  \int_{\mathcal{M}}  \frac{1}{e^2}\left(\lambda_1\lambda_2\lambda_3 \right)^\frac{4}{3} d\Omega_\mathcal{M}   =3\int_{\mathcal{M}}  \left( \frac{1}{e^{3/2}}\lambda_1\lambda_2\lambda_3 \right)^\frac{4}{3} d\Omega_\mathcal{M} \nonumber \\
&\geq&3 \frac{1}{(\mbox{vol}(\mathcal{M}))^{1/3}} \left( \int_{\mathcal{M}}  \frac{1}{e^{3/2}}\lambda_1\lambda_2\lambda_3 d\Omega_\mathcal{M}   \right)^\frac{4}{3} = 3 \frac{(\mbox{vol}(\Sigma))^{4/3}}{(\mbox{vol}(\mathcal{M}))^{1/3}}  \left( \left\langle\frac{1}{e^{3/2}} \right\rangle \right)^{4/3} |B|^{4/3} .
\eea
Restricting to the unit topological charge sector, we find that this bound is stronger than the former one if
\be
\mbox{vol}(\mathcal{M}) \leq 8 \frac{ \left\langle\frac{1}{e^{3/2}} \right\rangle^4}{ \left\langle\frac{f}{e} \right\rangle^3}\mbox{vol}(\Sigma).
\ee 

It is important that the numerical constants of these two bounds are given by independent target space averages. This opens the possibility that, having $\left\langle\frac{f}{e} \right\rangle$ finite, the other average, $\left\langle\frac{1}{e^{3/2}} \right\rangle$, may diverge, which would prevent the existence of the minimal dielectric Skyrmions on any finite volume domain. 

To see that such a case is realized, we consider the following choice of the coupling functions $e,f$ obeying the BPS constraint
\be
e=e_0\left( \frac{1}{2}\mbox{Tr }(1-U) \right)^\alpha = e_0(1-\cos \xi)^\alpha, \;\;\; f=f_0\left( \frac{1}{2}\mbox{Tr }(1-U) \right)^{1-\alpha}= f_0(1-\cos \xi)^{1-\alpha} \label{example}
\ee
where $\alpha \in \mathbb{R}$ while $e_0$ and $f_0$ are dimensional constants satisfying $r_0=1/(f_0e_0)$. The finiteness of the energy results from the finiteness of the $ \left\langle\frac{f}{e} \right\rangle$ average and requires that $\alpha < 5/4$. For all such $\alpha$ the models support a BPS $B=1$ hedgehog Skyrmion on the three dimensional Euclidean space $\mathcal{M}=\mathbb{R}^3$. On the other hand, finiteness of the average $ \left\langle\frac{1}{e^{3/2}} \right\rangle$ implies a different constraint on the parameter $\alpha$.
Specifically 
\be
 \left\langle\frac{1}{e^{3/2}} \right\rangle =\frac{2}{\pi} \frac{1}{e_0^{3/2}}\int_0^\pi \frac{\sin^2\xi }{(2\sin^2 \xi/2)^{\frac{3\alpha}{2}}} d\xi .
\ee
This integral converges if $\alpha < 1$. For $\alpha \geq 1$ the integral diverges and for a finite volume manifold $\mathcal{M}$ the r.h.s. of the bound is infinite. This means that there are no finite energy Skyrmions for such models (such a coupling function $e$) if the manifold $\mathcal{M}$ has a finite volume. 

The net result is that for $\alpha \in [1,5/4)$ the dielectric Skyrme model supports a unit charge Skyrmion on $\mathcal{M}=\mathbb{R}^3$ (which is in fact a BPS soliton saturating the pertinent topological bound), while it does not allow for finite energy topologically nontrivial solutions on any finite volume manifold. This means that these infinitely extended solitons {\it cannot} be enclosed in a finite volume. Formally it would require an infinite amount of energy to put this soliton in a finite volume. Hence, such a Skyrmion represents a completely incompressible three-dimensional matter.

%%%%%%%%%%%%%%%%%%%%%%%%%%%%%
\section{Incompressible perfect fluid solitons}
%%%%%%%%%%%%%%%%%%%%%%%%%%%%%
%%%%%%%%%%%%%%%%%%%%%%%%%%%%%
\subsection{Dielectric BPS Skyrme model}
%%%%%%%%%%%%%%%%%%%%%%%%%%%%%
To better understand the physics and mathematics of these incompressible solitons we will use another Skyrme type theory, i.e., the BPS Skyrme model \cite{adam2010skyrme}, again in its dielectric version. Note that the BPS Skyrme model contains the six-derivative term which provides the leading behavior at higher pressure/density \cite{Adam:2015lra}. This may have a nontrivial impact on properties of Skyrmionic matter in this regime resulting in a crystal-liquid phase transition in cores of neutron stars \cite{Adam:2014dqa}, \cite{Adam:2020yfv}. As we will see, the properties of incompressible Skyrmions in this model are quite analogous to the properties of the incompressible kinks. The dielectric BPS Skyrme model is defined by
\be
L^d_{60}=L^d_6+L^d_0= \int_{\mathcal{M}} \left(  g^2 \pi^4 \mathcal{B}_\mu^2 + \mathcal{U} \right) d\Omega_\mathcal{M}
\ee
where $\mathcal{U}$ is a non-derivative term (a potential) and $g$ is a target-space-dependent coupling function. This model is a BPS theory. Indeed, using the eigenvalues of the strain tensor  one can easily prove that the energy is bounded from below as
\be
E^d_{60}= \int_{\mathcal{M}} \left(  \frac{g^2}{4} \lambda_1^2\lambda_2^2\lambda_3^2 + \mathcal{U} \right) d\Omega_\mathcal{M} \geq \mbox{vol}(\Sigma)  \left\langle g\sqrt{\mathcal{U}} \right\rangle |B| .
\ee
The bound is saturated if and only if the corresponding Bogomolny equation is obeyed
\be
\frac{g}{2} \lambda_1\lambda_2\lambda_3 \pm \sqrt{\mathcal{U} }=0.
\ee
For a wide range of $g$, this equation admits topological solutions in any topological sector. The necessary condition is that the average $ \left\langle g\sqrt{\mathcal{U}} \right\rangle$ takes a finite value. As a consequence, the energy is a linear function of the topological charge which results in zero binding energies for all admissible coupling functions $g$ and potentials $\mathcal{U}$.

In addition, for any coupling function this model represents a perfect fluid \cite{adam2014thermodynamics}. Indeed, the energy-momentum tensor can be written in a perfect fluid form
\be
T^{\mu \nu} = (p+\rho) u^\mu u^\nu - p \eta^{\mu \nu}
\ee
where $\eta^{\mu \nu}$ is the Minkowski metric. In the static case, the four velocity is $u^\mu=(1,0,0,0)$, and the energy density and pressure are, respectively 
\be
\rho = g^2\pi^4 \mathcal{B}_\mu^2 + \mathcal{U}, \;\;\; p = g^2\pi^4 \mathcal{B}_\mu^2 - \mathcal{U}.
\ee
The conservation of the energy-momentum tensor, $\partial_\mu T^{\mu \nu}=0$, implies that the pressure $p$ must be a constant. 
 
Furthermore, the static energy functional is invariant under the volume-preserving diffeomorphisms of the base space. This means that a BPS soliton with a given topological charge can have an arbitrary shape provided that its volume remains unchanged also locally. This again corresponds with the symmetries of a perfect fluid with no surface (tension) term. 

Let us again consider a finite volume base space $\mathcal{M}$. Then, 
\bea
E^d_{60} &=& \int_{\mathcal{M}} \left(  \frac{g^2}{4} \lambda_1^2\lambda_2^2\lambda_3^2 + \mathcal{U} \right) d\Omega_\mathcal{M} \geq  \int_{\mathcal{M}} \frac{g^2}{4} \lambda_1^2\lambda_2^2\lambda_3^2 \, d\Omega_\mathcal{M} \\
&\geq&\frac{1}{\mbox{vol}(\mathcal{M})}   \left(  \int_{\mathcal{M}} \frac{g}{2} \lambda_1\lambda_2\lambda_3 \, d\Omega_\mathcal{M} \right)^2 = \frac{1}{8\pi^2} \frac{\mbox{vol}(\Sigma)}{\mbox{vol}(\mathcal{M})}   \left( \left\langle g \right\rangle \right)^2 B^2.
\eea
As in the case of the dielectric extension of the minimal Skyrme model $L_{24}^d$, the average appearing in the finite volume bound, $ \left\langle g \right\rangle$, is independent of the average in the general bound, $ \left\langle g \sqrt{\mathcal{U}} \right\rangle$. Therefore we can, again, find a situation where $ \left\langle g \right\rangle = \infty$ while $ \left\langle g \sqrt{\mathcal{U}} \right\rangle$ takes a finite value. As a consequence, a (BPS) soliton with any topological charge exists on $\mathbb{R}^3$ but not in a finite base space, which gives rise to completely incompressible topological solitons.  

As a particular example we consider the following choice of the coupling function $g$ and the potential
\be
g=g_0 \eta^{-1}, \;\;\; \mathcal{U}=\mu^2 \eta^2 \label{example}
\ee
where $g_0$ and $\mu$ are dimensional constants while $\eta$ is a new target space coordinate related to the usual $\xi$ (which is further related to $\mbox{Tr } U$) as
\be
\eta = \frac{1}{2} \left(\xi - \frac{1}{2} \sin 2\xi \right) .
\ee
In the unit topological charge sector we can again assume the hedgehog ansatz (for higher values of the topological charge one has to use the axially symmetric ansatz). Then the Bogomolny equation on $\mathbb{R}^3$ gives
\be
\frac{1}{2r^2} g \sin^2\xi  \xi_r = -\sqrt{\mathcal{U}} \;\; \Rightarrow \;\; \frac{g_0}{2r^2}  \frac{\eta_r}{\eta} = -\mu \eta
\ee
with a simple solution
\be
\eta = \frac{1}{\frac{2\mu}{3g_0} r^3+ \frac{2}{\pi}} 
\ee
interpolating between $\eta (r=0)=\pi/2$ (hence $\xi(r=0)=\pi$) and $\eta (r \to \infty)=0$ (hence $\xi(r\to \infty)=0$). However, for this coupling function $g$ the average $ \left\langle g \right\rangle $ is logarithmically divergent. Thus this soliton cannot be put in a finite volume space. 
%%%%%%%%%%%%%%%%%%%%%%%%%%%%%
\subsection{The BPS Skyrme model and pressure}
%%%%%%%%%%%%%%%%%%%%%%%%%%%%%
As we have already shown, the BPS Skyrme model, also in its dielectric version, is a perfect fluid theory. As a consequence of this fact, the pressure $p$ appears as a field theoretical parameter in the first integral of the static field equations. In fact, the full static second order partial differential equation is integrable to a constant pressure equation \cite{adam2014thermodynamics}
\be
p = g^2\pi^4 \mathcal{B}_\mu^2 - \mathcal{U}. \label{const-pres}
\ee
Note that for $p=0$ we recover the Bogomolny equation. Hence the BPS solitons obey the zero pressure equation.

Let us briefly summarize the results known for the usual BPS Skyrme model where $g$ is simply a constant. When the pressure vanishes, the Bogomolny equation gives rise to topologically nontrivial solutions for any reasonable one-vacuum potential $\mathcal{U}$ where the vacuum is chosen to be $U=\mathbb{I}$. We consider a large class of potentials that depend on $\mbox{Tr } U$ i.e., on $\xi$. The behavior of the potential close to the vacuum,  $\mathcal{U} \sim \xi^a$, where $a >0$, determines the qualitative type of the soliton. If $a \in (0,6)$ then the resulting solitons are compactons which differ from the vacuum $\xi=0$ only in a finite region of space. For $a\geq 6$, we obtain the usual infinitely extended Skyrmions with exponential or power-like tails. Obviously, the geometric volume $V$ of compactons is finite while in the case of usual solitons it takes an infinite value. 

When a non-zero pressure is applied, the solitons are squeezed and their geometric volume is reduced \cite{adam2014thermodynamics}
\be
V(p)=\pi^2 |B| g   \left\langle \frac{1}{\sqrt{\mathcal{U} +p}} \right\rangle .
\ee 
For any positive pressure $\sqrt{\mathcal{U} + p} > \sqrt{\mathcal{U}}$ which implies that $V(p)<V(p=0)$. In addition, as $\mathcal{U} + p >0$, the volume is always finite if a non-zero pressure is applied, $V(p) < \infty$. This follows from the fact that the average is an integral over the three dimensional unit sphere with a finite volume. Note that only positive pressure is admissible. Indeed, for $p<0$ the constant pressure equation does not allow to approach the vacuum where $\mathcal{U}= 0$.  Therefore, any infinitely extended BPS soliton can be squeezed to a finite volume by imposing a non-zero (finite) pressure. 

When the soliton is compressed, its energy grows as  \cite{adam2014thermodynamics}
\be
E_{60}(p)=\pi^2 g |B|  \left\langle \frac{2\mathcal{U} +p}{\sqrt{\mathcal{U} +p}} \right\rangle .
\ee
One can also verify that the geometric volume is the proper thermodynamical volume satisfying the standard thermodynamical relation 
\be
p=-\left( \frac{\partial E_{60} }{\partial V} \right)_B . \label{thermo}
\ee 
%%%%%%%%%%%%%%%%%%%%%%%%%%%%%
\subsection{Dielectric BPS Skyrme model and pressure}
%%%%%%%%%%%%%%%%%%%%%%%%%%%%%
Now we turn back to our example (\ref{example}) and consider the constant pressure equation (\ref{const-pres})
\be
\frac{g^2_0}{4r^4}  \frac{\eta^2_r}{\eta^2} - \mu^2 \eta^2=p .
\ee
Again $p\geq 0$. The corresponding formal solutions read
\be
\eta= \frac{\sqrt{p}}{\mu \sinh \frac{\sqrt{p}}{\mu} \left( \frac{2 \mu r^3}{3g_0} + z_0\right) } \label{formal}
\ee
where $z_0$ obeys 
\be
 \frac{\sqrt{p}}{\mu \sinh \frac{\sqrt{p}}{\mu}  z_0 } = \frac{\pi}{2}.
\ee
This guarantees that the topologically nontrivial boundary conditions $\eta(r=0)=\pi/2$ and $\eta(r\to \infty)=0$ are satisfied. 

Surprisingly, we see that for any finite pressure the skyrmions are still infinitely extended. Hence, an addition of a non-zero pressure does not reduce the geometrical volume of the soliton. It becomes better localized but still extends to spatial infinity. Specifically, the Bogomolny equation allows to relate the base space integral to a target space integral using the fact that the base space volume form can be expressed as the pullback of a target space three-form via 
\be
d\Omega_\mathcal{M} = \left( \frac{g\pi^2}{\sqrt{\mathcal{U} +p}} \frac{d\Omega_{\Sigma}}{2\pi^2} \right)^*
\ee
where now $g$ is a target space function. Then, the volume of the soliton reads
\be
V(p)=\int d\Omega_\mathcal{M} = \pi^2 |B|    \left\langle \frac{g}{\sqrt{\mathcal{U} +p}} \right\rangle .
\ee 
This explains the incompressible nature of these solitons and why they cannot be put on a finite manifold. 

In addition, solutions (\ref{formal}) are in fact formal, infinite energy solution. Indeed, the energy of the solitons can also be computed as a target space integral
\be
E^d_{60}(p)=\pi^2 |B|  \left\langle g \frac{2\mathcal{U} +p}{\sqrt{\mathcal{U} +p}} \right\rangle .
\ee

Now it is clearly seen that at any non-zero pressure the volume of the soliton and its energy are decided by the behavior of the coupling function $g$. If $ \left\langle g \right\rangle $ diverges, then the volume and energy are infinite for any nonzero pressure. At $p=0$, the divergency of $ \left\langle g \right\rangle $ in the energy integral can be softened or even canceled by the potential. This finally guaranties the existence of the BPS (zero pressure) solitons on $\mathbb{R}^3$. 

%%%%%%%%%%%%%%%%%%%%%%%%%%%%%
\section{Summary and applications}
%%%%%%%%%%%%%%%%%%%%%%%%%%%%%
In the present work, we have identified a new class of topological solitons, in addition to the usual infinitely extended solitons and the compactons. They can exist on a base space of infinite volume, here the Euclidean space $\mathbb{R}^n$, where they are infinitely extended. However, in contrast to the usual solitons and compactons, they cannot be closed in a finite volume. Basically, finite energy topologically nontrivial solutions transform into infinite energy configurations if put on a finite volume manifold. Therefore, they may be interpreted as totally {\it incompressible solitons}. 

A related property, found in the case of models which correspond to perfect fluid theories (kinks in (1+1) dimensions and Skyrmions in the dielectric BPS Skyrme model in (3+1) dimensions), is that these incompressible solitons are resistant to any pressure. In fact, formal topologically nontrivial solutions with a non-zero pressure are found but they have infinite energy. It means that it requires an infinite amount of energy to increase pressure in such a solitonic matter. Furthermore, solutions with non-zero pressure always extend to infinity although they are better localized than in the zero-pressure case. 

These properties, i.e., the nonexistence of finite volume and finite pressure solutions, make such solitons qualitatively very distinct from the typical infinitely extended solitons and compactons, allowing us to define a new third class of {\it incompressible} solitons. 
It should be underlined that the family of incompressible solitons is quite general. These objects are not confined to one specific field theory. On the contrary, they exist in various theoretical set-ups (in various dimensions) describing different physical situations. 

Interestingly, an example of such incompressible solitons is provided by a family of standard scalar field theories in (1+1) dimensions known as runaway (vacuumless) potentials. This kind of models found some applications in the context of the so-called quintessence models \cite{Peebles1988, Ratra1988, Caldwell1998} and spatial and/or time variation of fundamental coupling constants \cite{Chiba:2011en}, \cite{Bamba:2011nm}. It would be interesting to identify a physical imprint of the incompressibility of the topological defects in these physical models. In any case, our finding may allow us to look at these models from a different point of view, providing a new and deeper insight or a reinterpretation of previously known results. 

A runaway potential has been also considered in a version of the Abelian Higgs model \cite{Marques:2018xow}. One can expect that the obtained vortices provide another example of incompressible solitons.

Looking from a wider perspective,  incompressible solitons can be physically relevant, as fluids are virtually incompressible in a first approximation. So, it could be interesting to apply this class of solutions to understand the impact of the compressibility on the dynamics of topological solitons, detecting phenomena which are strongly affected by a transition from incompressible to compressible matter. Probably, to simplify the situation one should start with incompressible kinks in runaway field theories. In fact, a kink-antikink scattering in a family of models, interpolating between a runaway potential $V=1/\cosh^2 \phi$ and standard two vacuum potentials, has been investigated \cite{Simas2017}. Unfortunately, the paper focuses mainly on the standard two vacua potential, and very little results concerning the runaway case have been reported. We will address the problem of the interaction of incompressible kinks in a forthcoming paper.

Importantly, it is the degeneracy pressure (exclusion principle pressure) which makes physical liquids and solids quite incompress­ible un­der nor­mal pres­sures. Hence, it is tempting to relate the incompressible solitons to the notion of the degeneracy pressure. This can be relevant e.g., for the description of atomic nuclei in terms of Skyrmions. Undoubtedly, more research is needed. 

Finally, incompressible solitons are also relevant for gauge theories. Indeed, in the large-$N_c$ limit the instanton liquid becomes incompressible \cite{nowak1989instantons}. 

%%%%%%%%%%%%%%%%%%%%%%%%%%%
\section*{Acknowledgements}
%%%%%%%%%%%%%%%%%%%%%%%%%%%%
CA and AW acknowledge financial support from the Ministry of Education, Culture and Sports, Spain (Grant No. FPA2017-83814-P), the Xunta de Galicia (Grant No. INCITE09.296.035PR and Conselleria de Educacion), the Spanish Consolider-Ingenio 2010 Programme CPAN (CSD2007-00042), Maria de Maetzu Unit of Excellence MDM- 2016-0692, and FEDER.
KO, TR and AW were supported by the Polish National Science Centre, grant NCN 2019/35/B/ST2/00059. CN is supported by the INFN grant 19292/2017 (MMNLP) "Integrable Models and Their Applications to Classical and Quantum Problems". AW thanks Maciej Nowak for discussion.

\bibliographystyle{JHEP}
\bibliography{incompress}

\end{document}